\documentclass[12pt]{article}
\pdfoutput=1
\usepackage{graphicx}
\usepackage{times}

\title{Fiber-Optical Analogue of the Event Horizon}
\author{Thomas G.\ Philbin$^{1,2}$,
Chris Kuklewicz$^{1}$,
Scott Robertson$^{1}$,
Stephen Hill$^{1}$,\\
Friedrich K\"onig$^{1}$,
and
Ulf Leonhardt$^{1\ast}$\\
\normalsize{$^1$School of Physics and Astronomy,}\\
\normalsize{University of St Andrews,North Haugh, St Andrews, Fife, KY16 9SS, UK}\\
\normalsize{$^2$Max Planck Research Group of Optics, Information and Photonics,}\\
\normalsize{G\"unther-Scharowsky-Str.\ 1, Bau 24, D-91058 Erlangen, Germany}\\
\\
\normalsize{$^\ast$To whom correspondence should be addressed; E-mail:  ulf@st-andrews.ac.uk.}
}

\date{}

\begin{document}

\maketitle

\begin{abstract}
The physics at the event horizon resembles the
behavior of waves in moving media. 
Horizons are formed where the local speed of the medium 
exceeds the wave velocity.
We use ultrashort pulses in microstructured optical fibers
to demonstrate the formation 
of an artificial event horizon in optics.
We observed a classical optical effect,
the blue-shifting of light at a white-hole horizon. 
We also show by theoretical calculations that such a system is 
capable of probing the quantum effects of horizons, 
in particular Hawking radiation.
\end{abstract}

\newpage

Laboratory analogues of black holes 
\cite{Artificial,Volovik,SUbook}
are inspired by
a simple and intuitive idea \cite{Unruh}:
the space-time geometry of a black hole resembles a river 
\cite{Jacobson,Rousseaux}, 
a moving medium flowing towards a waterfall,
the singularity.
Imagine that the river carries waves 
propagating against the current with speed $c'$.
The waves play the role of light
where $c'$ represents $c$, the speed of light in vacuum.
Suppose that the closer
the river gets to the waterfall the faster it flows
and that at some point the speed of the river exceeds $c'$.
Clearly, beyond this point
waves can no longer propagate upstream.
The point of no return corresponds to the horizon of the black hole.
Imagine another situation:
a fast river flowing out into the sea, getting slower.
Waves cannot enter the river beyond the point
where the flow speed exceeds the wave velocity;
the river resembles an object that nothing can enter,
a white hole.

Nothing, not even light, can escape 
from a gravitational black hole.
Yet according to quantum physics,
the black hole is not entirely black, 
but emits waves in thermal equilibrium 
\cite{Hawking,BD,Brout}.
The waves consist of correlated pairs of quanta, 
one originates from inside and the other from outside the horizon.
Seen from one side of the horizon,
the gravitational black hole acts as  
a thermal  black-body radiator
sending out Hawking radiation 
\cite{Hawking,BD,Brout}.
The effective temperature depends on 
the surface gravity \cite{Hawking,BD,Brout}
that, in the analogue model, corresponds to
the flow-velocity gradient at the horizon
\cite{Artificial,Volovik,SUbook,Unruh,Jacobson}.

The Hawking temperature of typical black holes 
lies far below the temperature of the cosmic microwave background,
so an observation of Hawking radiation in astrophysics 
seems unlikely. 
However, laboratory demonstrations of analogues 
of Hawking radiation could be feasible.
One type of recent proposal \cite{Garay,GFKL,Giovanazzi}
suggests the use of ultracold quantum gases such as 
alkali Bose-Einstein condensates 
or ultracold alkali Fermions \cite{Giovanazzi}.
When a condensate in a waveguide is pushed over a potential barrier 
it may exceed the speed of sound (typically a few mm/s) 
and is calculated to 
generate a Hawking temperature of about $10\mbox{nK}$ 
\cite{GFKL}. 
Helium-3 offers a multitude of analogues between quantum fluids
and the standard model, including Einsteinian gravity
\cite{Volovik}.
For example, 
the analogy between gravity and surface waves in fluids
\cite{SU0} has inspired ideas for artificial event horizons
at the interface between two sliding superfluid phases
\cite{Volbrane},
but, so far, none of the quantum features of horizons 
has been measured in Helium-3.
Proposals for optical black holes \cite{LPUsch,Leo} 
have relied on slowing down light  \cite{Milonni}
such that it matches
the speed of the medium \cite{LPUsch} 
or on bringing light
to a complete standstill \cite{Leo}, 
but in these cases absorption 
may pose a severe problem near the horizon 
where the spectral transparency window \cite{Milonni} vanishes. 

On the other hand, 
is it necessary to physically move a medium for establishing a horizon?  
What really matters are only the effective properties of the medium. 
If they change, for example as a propagating front, 
but the medium itself remains at rest, 
the situation is essentially indistinguishable from a moving medium. 
Such ideas have been discussed for moving solitons and domain walls 
\cite{JV} in superfluid Helium-3 \cite{Volovik}
and more recently for microwave transmission lines 
with variable capacity \cite{SU}, 
but they have remained impractical so far. 

Here we report the first experimental observation
of the classical optical effects of horizons, 
the blue-shifting of light at a white-hole horizon, 
and show theoretically
that our scheme combines several promising features
for demonstrating quantum Hawking radiation 
in the optical domain.
Our idea, illustrated in Fig.\ 1,
is based on the nonlinear optics
of ultrashort light pulses in optical fibers \cite{Agrawal}
where we exploit the remarkable control
of the nonlinearity, birefringence and dispersion
in microstructured fibers \cite{Russell,Reeves}.

Using a Titanium-Sapphire laser 
we create $70\mathrm{fs}$-long
non-dispersive pulses (solitons)
at $803\mathrm{nm}$ carrier wavelength
and $80\mathrm{MHz}$ repetition rate
inside $1.5\mathrm{m}$ of microstructured fiber
(NL-PM-750B from Crystal Fibre A/S).
Each pulse modifies the optical properties of the fiber
due to the Kerr effect \cite{Agrawal}:
the effective refractive index of the fiber, $n_0$,
gains an additional contribution ${\delta n}$
that is proportional to the instantaneous pulse intensity
$I$ at position $z$ and time $t$,
\begin{equation}
n = n_0 + {\delta n}, \quad {\delta n} \propto I(z,t)\,.
\label{eq:medium}
\end{equation}
This contribution to the refractive index $n$
moves with the pulse.
The pulse thus establishes a moving medium,
although nothing material is moving. 
This effective medium naturally moves 
at the speed of light in the fiber,
because it is made by light itself.

We also launch a continuous wave of light, a probe, that
follows the pulse with slightly higher group velocity, 
attempting to overtake it.
In order to distinguish the probe from the pulse,
it oscillates at a significantly different frequency $\omega$.
Our probe-light laser is tunable over wavelengths $2\pi c/\omega$
from $1460\mathrm{nm}$ to $1540\mathrm{nm}$.
While approaching the pulse,
the Kerr contribution ${\delta n}$ slows down the probe
until the probe's group velocity reaches the speed of the pulse.
The trailing end of the pulse establishes a white-hole 
horizon, an object that light cannot enter, 
unless it tunnels through the pulse.
Conversely, the front end creates a black-hole horizon
for probe light that is slower than the pulse. 
As ${\delta n}$ is small,
the initial group velocity
of the probe should be sufficiently close 
to the speed of the pulse.
In microstructured fibers \cite{Russell}
the group-velocity dispersion of light is engineered
by arrangements of air holes
(sub-$\mu\mathrm{m}$ wide hollow cylinders along the fiber).
We selected a fiber where
the group velocity of pulses near the $800\mathrm{nm}$
carrier wavelength of mode-locked Ti:Sapphire lasers
matches the group velocities of probes 
in the infrared telecommunication band 
around $1500\mathrm{nm}$,
whereas standard optical fibers \cite{Agrawal}
do not have this property.

At the horizon of an astrophysical black hole light freezes, 
reaching wavelengths shorter than 
the Planck scale where the physics is unknown.
(The Planck length is given by $\sqrt{2\pi\hbar G /c^3}$
where $G$ is the gravitational constant.)
Some elusive Trans-Planckian mechanism must regularize 
this behavior \cite{tHooft,JacobsonTrans}.
In our case, the fiber-optical analogue of 
Trans-Planckian physics is known and simple ---
it is contained in the frequency-dependance of the
refractive index $n$, the dispersion of the fiber:
at the trailing end of the pulse
the incoming probe modes are compressed,
oscillating with increasing frequency;
they are blue-shifted.
In turn, the dispersion limits the frequency shifting by
tuning the probe out of the horizon.
In the case of normal group-velocity dispersion
the blue-shifted light falls behind.
At the black-hole horizon the reverse occurs: 
a probe slower than the pulse is red-shifted 
and then moves ahead of the pulse.

Figure 2 shows the difference
in the spectrum of the probe light
--- incident with $\omega_1$ ---
with and without the pulses,
clearly displaying a blue-shifted peak at $\omega_2$.
To quantitatively describe this effect, 
we consider the frequency $\omega'$ in a co-moving frame
that is connected to the laboratory-frame frequency $\omega$
by the Doppler formula
\begin{equation}
\omega' = \left(1-\frac{nu}{c}\right)\omega \,.
\label{eq:doppler}
\end{equation}
For a stable pulse, $\omega'$ is a conserved quantity,
whereas $\omega$ follows the contours of fixed $\omega'$
when ${\delta n}$ varies with the 
intensity profile of the pulse (Fig.\ 3).
For sufficiently large ${\delta n}$,
the frequency $\omega$ completes an arch 
from the initial $\omega_1$ to the final $\omega_2$;
it is blue-shifted by the white-hole horizon. 
At a black-hole horizon, the arch is traced the other way round
from $\omega_2$ to $\omega_1$.
For the frequency at the center of the arches
an infinitesimal ${\delta n}$ is sufficient to cause 
a frequency shift; at this frequency 
the group velocity of the probe
matches the group-velocity of the pulse.
Figure 2 shows that
both the blue-shifted and probe light are spectrally broadened.
These features are easily explained:
the horizon acts only during the time while 
probe and pulse propagate in the fiber,
where only a finite fraction of the probe is frequency-shifted,
forming a blue-shifted pulse and also
a gap in the probe light, a negative pulse;
these pulses have a characteristic spectrum
with a width that is inversely proportional
to the fiber length.
We compared the measured spectra
with the theory of light propagation 
in the presence of horizons,
and found very good agreement
\cite{SOM}.

Imagine instead of a single probe a set of probe modes.
The modes should be sufficiently weakly excited such that 
they do not interact with each other via the Kerr effect,
but they experience the cross Kerr effect of the pulse,
the presence of the medium (\ref{eq:medium}) 
moving with the velocity $u$.
The modes constitute a quantum field of light
in a moving medium \cite{SOM,LeoReview}.
Classical light is a real electromagnetic wave.
So, according to Fourier analysis, 
any amplitude oscillating as $\exp(-i\omega\tau)$
at a positive angular frequency $\omega$
must be accompanied by  
the complex conjugate amplitude at $-\omega$.
In quantum field theory \cite{BD,Brout,SOM},
the positive-frequency modes correspond to
the annihilation and the negative-frequency modes
to the creation operators \cite{LeoReview}.
Processes that mix positive and negative frequencies 
in the laboratory frame (in the glass of the fiber) 
thus create observable light quanta.

In the near ultraviolet around $300\mathrm{nm}$,
the dispersion of microstructured fibers  \cite{Russell}
is dominated by the bare dispersion of glass where
$n_0$ rapidly grows with frequency \cite{Agrawal}, 
exceeding $c/u$.
For such ultraviolet modes, 
the medium moves at superluminal speed.
We see from the Doppler formula (\ref{eq:doppler})
that these superluminal modes oscillate with negative
frequencies $\omega'$ in the co-moving frame
for positive frequencies $\omega$ in the laboratory frame,
and vice versa.
Moreover, 
each subluminal mode with positive $\omega$
has a superluminal partner
oscillating at the same co-moving frequency $\omega'$,
but with negative laboratory frequency.
The pulse does not change $\omega'$,
but it may partially convert sub- and superluminal
partner modes into each other,
thus creating photons \cite{BD,Brout}.
Even if all the modes are initially in their vacuum states,
the horizon spontaneously creates photon pairs.
This process represents 
the optical analogue of Hawking radiation \cite{Hawking}
and it occurs at both the black-hole 
and white-hole horizon \cite{SOM}.
Photons with positive $\omega'$ 
correspond to the particles 
created at the outside of the black hole \cite{BD,Brout},
while the negative-frequency photons represent 
their partners beyond the horizon.
In our case,
the photon pairs are distinguishable from the intense pulse,
because their polarization can be orthogonal to the pulse
and their frequencies differ from the carrier frequency by an octave.
Furthermore, one can discriminate
the Hawking effect from other nonlinear optical processes,
such as Four Wave Mixing,
because it is not subject to their phase-matching conditions 
\cite{Agrawal}.  
Moreover, in addition to observing Hawking radiation per se,
one could detect the correlations of the Hawking partners ---
a feat that is impossible in astrophysics,
because there the partner particles are lost beyond the horizon
of the black hole.

In order to give a quantitative argument for the Hawking effect
in optical fibers,
we describe the propagation distance $z$ 
in terms of the time $\zeta$ 
it takes for the pulse to traverse it, $\zeta=z/u$,
and introduce the retarded time $\tau=t-z/u$.
The phase $\varphi$
of each mode evolves as 
\begin{equation}
\varphi = - \int \big(\omega\mathrm{d}\tau + 
\omega'\mathrm{d}\zeta \big) \,.
\label{eq:phase}
\end{equation}
We assume that the mode conversion occurs in a narrow interval
of retarded time $\tau$ near a horizon around $\tau=0$,
where we linearize ${\delta n}$ in $\tau$ such that
\begin{equation}
1-\frac{nu}{c} = \alpha'\tau \,.
\label{eq:linear}
\end{equation}
We obtain from the phase integral (\ref{eq:phase})
and the Doppler formula (\ref{eq:doppler})
the characteristic logarithmic phase at a horizon
\cite{BD,Brout}.
We use the standard result \cite{BD,Brout,SOM}:
Hawking radiation is Planck-distributed with the 
temperature
\begin{equation}
k_B T' = \frac{\hbar \alpha'}{2\pi} \,,
\end{equation}
where $k_B$ denotes Boltzmann's constant.
For evaluating $\alpha'$ we consider ${\delta n}$ at $\tau=0$, where
\begin{equation}
\alpha' = \left.-\frac{u}{c}\,\frac{\partial n}{\partial\tau}\right|_0
= \left.-\frac{u}{c}\,\frac{\partial {\delta n}}{\partial\tau}\right|_0
\,.
\end{equation}
Note that $T'$ denotes the Hawking temperature 
in the co-moving frame, defined with respect to the
Doppler-shifted frequencies $\omega'$,
a temperature that is characterized by the Doppler-shifted
Hawking frequency $\alpha'$
in regions away from the pulse.
We use the Doppler formula (\ref{eq:doppler})
with the refractive index (\ref{eq:medium})
and the linearization (\ref{eq:linear})
taken at $\tau=0$, and obtain
\begin{equation}
\alpha' = \left(1-n_0\frac{u}{c}\right)\alpha = 
\left.\frac{u}{c}\, {{\delta n}} \right|_0 \alpha
\,.
\end{equation}
Consequently, the Hawking temperature $T$ 
in the laboratory frame is
\begin{equation}
k_B T = \frac{\hbar \alpha}{2\pi} \,,\quad
\alpha = 
-\left.\frac{1}{{\delta n}}\, \frac{\partial{\delta n}}{\partial\tau}\right|_0
\,.
\label{eq:hawking}
\end{equation}
As $T$ does not depend on the magnitude of ${\delta n}$,
even the typically very small refractive-index
variations of nonlinear fiber optics \cite{Agrawal}
may lead to a substantial Hawking temperature
when ${\delta n}$ varies on the scale of an optical wavelength.
This is achievable with few-cycle optical pulses 
\cite{Kartner,Krausz}.

Note that the Kerr nonlinearity \cite{Agrawal}
not only influences the probe modes, 
but the pulse as well \cite{Agrawal}.
This self Kerr effect shapes the pulse 
while it propagates in the fiber. 
Regions of high intensity lag behind, 
because for them the effective refractive index is increased. 
The black-hole horizon at the front is stretched,
but the trailing edge becomes extremely steep,
infinitely steep in theory \cite{Agrawal}:
the pulse develops an optical shock \cite{Agrawal}.
The steep white-hole horizon 
will dominate the Hawking effect of the pulse.  
In practice, dispersion combined with 
other nonlinear optical processes in the fiber,
in particular Stimulated Raman Scattering \cite{Agrawal},
limit the optical shock.
Assuming that the steepness at the shock front is 
comparable to twice the frequency of the pulse carrier,
$8\times10^{14}\mathrm{Hz}$,
the Hawking temperature (\ref{eq:hawking})
reaches $10^3\mbox{K}$,
many orders of magnitude higher than
condensed-matter analogues of the event horizon
\cite{Garay,GFKL,Giovanazzi,JV}.

Our scheme thus solves two problems at once in a natural way: 
how to let an effective medium move at superluminal speed 
and how to generate a steep velocity profile at the horizon;  
the various aspects of the physics of artificial black holes 
conspire together, in contrast to most other proposals 
\cite{Artificial,Volovik,SUbook,Unruh,Garay,GFKL,Giovanazzi,SU0,Volbrane,LPUsch,Leo}.


\newpage

\begin{figure}[t]
\begin{center}
\includegraphics[width=30.0pc]{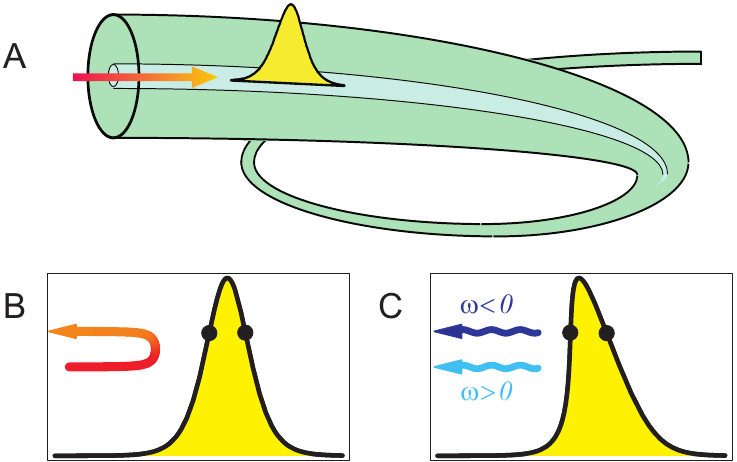}
\caption{
\small{
Fiber-optical horizons.
A a light pulse in a fiber slows down infrared probe light
attempting to overtake it. 
The diagrams below are in the co-moving frame of the pulse.
B Classical horizons.
The probe is slowed down by the pulse
until its group velocity matches the pulse speed 
at the points indicated in the figure,
establishing a white-hole horizon at the back 
and a black-hole horizon at the front of the pulse.
The probe light is blue-shifted at the white hole
until the optical dispersion releases it from the horizon.
C Quantum pairs.
Even if no probe light is incident,
the horizon emits photon pairs
corresponding to
waves of positive frequencies
from the outside of the horizon paired with
waves at negative frequencies
from beyond the horizon.
An optical shock has steepened the pulse edge,
increasing the luminosity of the white hole.
}
\label{fig:scheme}}
\end{center}
\end{figure}

\newpage

\begin{figure}[h]
\begin{center}
\includegraphics[width=30.0pc]{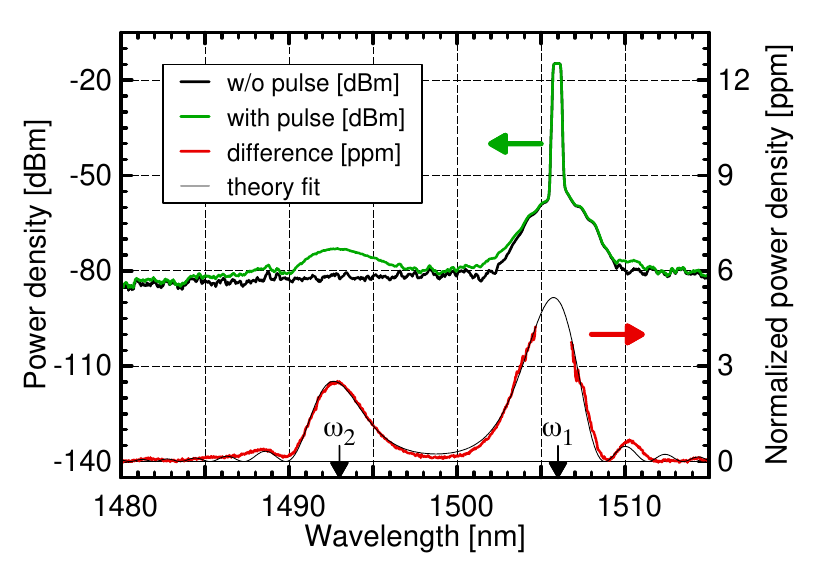}
\caption{
\small{
Measurement of blue shifting
at a white-hole horizon. 
The black curve shows the power spectrum 
of probe light that has not interacted with the pulses,
while the green curve displays the result of the interaction;
both curves are represented on a logarithmic scale.
The difference between the spectra on a linear scale, 
shown in red, exhibits a 
characteristic peak around the blue-shifted wavelength 
($\omega_2$)
and another peak around the spectral line of the probe laser
($\omega_1$)
due to a gap in the probe light;
both features indicate the presence of a horizon.
}
\label{fig:spectrum}}
\end{center}
\end{figure}

\newpage

\begin{figure}[h]
\begin{center}
\includegraphics[width=30.0pc]{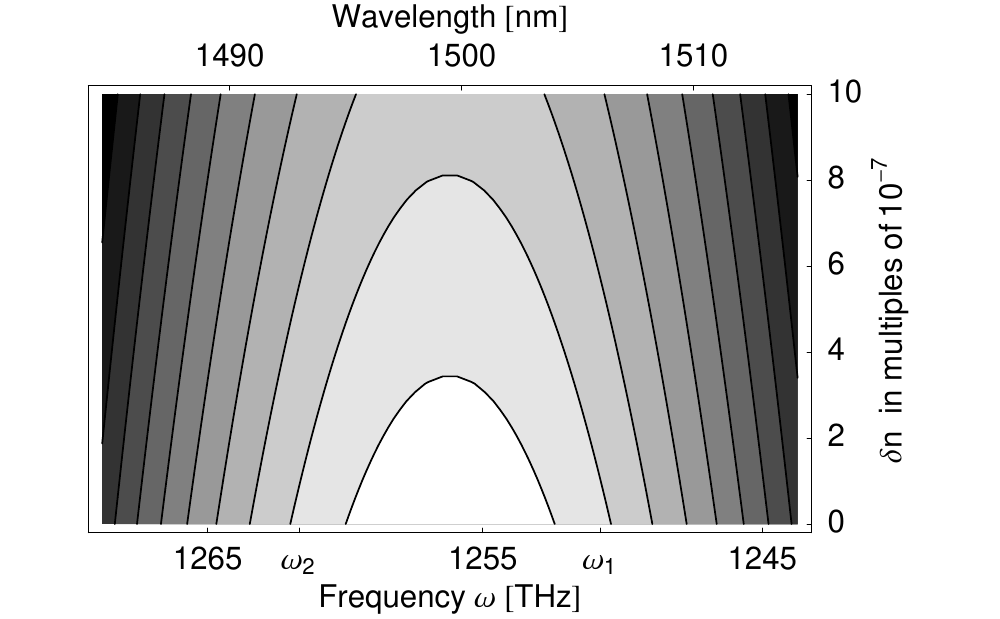}
\caption{
\small{
Doppler contours.
The Doppler-shifted frequency $\omega'$ of the probe
is a conserved quantity.
The pulse shifts the laboratory frequency $\omega$ 
along the contour lines of
$\omega'$ as a function of the 
instantaneous $\delta n$;
the same applies to the wavelength 
$\lambda=2\pi c/\omega$.
If the pulse is sufficiently intense 
such that $\delta n$ reaches the top of a contour, 
the probe light  
completes an arch on the diagram
while leaving the pulse;
it is red- or blue-shifted,
depending on its initial frequency.
}
\label{fig:contours}}
\end{center}
\end{figure}

\end{document}